\documentclass[twocolumn,showpacs,aps,prl,
groupedaddress,amssymb,amsmath,nobalancelastpage]{revtex4}

\usepackage{graphicx}
\usepackage{longtable}

\begin{document}

\title{Asymmetric response of a jammed plastic bead raft}

\author{Michael Twardos}
\author{Michael Dennin}
\affiliation{Department of Physics and Astronomy, University of
California at Irvine, Irvine, California 92697-4575}

\date{\today}

\begin{abstract}

Fluctuation-dissipation relations have received significant
attention as a potential method for defining an effective
temperature in nonequilibrium systems. The successful development
of an effective temperature would be an important step in the
application of statistical mechanics principles to systems driven
far from equilibrium. Many of the systems of interest are
sufficiently dense that they are close to the jamming transition,
a point at which interesting correlations develop. Here we study
the response function in a driven system of plastic beads as a
function of the density in order to elucidate the impact of the
jamming transition on the use of fluctuation-dissipation
relations. The focus is on measuring the response function for
applied shear stress. We find that even when the amplitude of the
applied stress leads to a linear response in the strain, the time
scale of the response is dependent on the direction of the applied
stress.

\end{abstract}


\maketitle

The identification of an ``effective temperature'' for systems
driven far from equilibrium that is analogous to a true
thermodynamic temperature would represent a major step toward
developing a general theory of nonequilibrium behavior. One of the
reasons that the concept of an ``effective temperature'' is so
attractive is the centrality of real temperature in thermodynamics
and statistical physics. At its most basic, temperature defines
when two systems are in thermodynamic equilibrium and determines
the direction of heat flow, if any, for systems in contact.
Equipartition provides a connection between temperature and
average quantities. Temperature enters the relation between
thermodynamic quantities, such as specific heat and derivatives of
free energies. Einstein-type relations relate various transport
coefficients, such as diffusion constants and viscosities, through
the temperature. Finally, in linear response theory, temperature
enters the relationship between time (or frequency) dependent
response functions and correlation functions. The challenge for
studies of nonequilibrium systems is the determination of which,
if any, of the above uses of temperature is meaningful as an
effective temperature. To be a useful concept, a minimum
expectation is that multiple definitions of effective temperature
agree, and that there is an understanding of why some definitions
agree and others do not.

The use of effective temperatures in driven systems has a
relatively diverse history. For example, granular flows are often
characterized by a ``granular'' temperature that is based on the
average kinetic energy of the particles. Another approach can be
described as a ``configurational'' temperature. In these studies,
(usually with dense granular matter) the ``atoms'' are grains, and
the phase space is the set of static mechanical configurations in
which they can be arranged. By considering all possible jammed
configurations, entropy as a function of density is calculated and
used to define an effective temperature \cite{JC99,DG04,EBM05}.
Another approach has been to focus on fluctuations in the system
as a means of defining a temperature. In this case, the
probability distribution of an appropriate macroscopic variable,
such as the power or stress, is studied, and the temperature is
related to various measures of this distribution. This approach
has been used in turbulent fluid systems
\cite{CGHLPC04,GGK01,TG03} and driven complex fluids
\cite{GA03,FM04}.

In this paper, we focus on definitions of effective temperature
based on the relationship between linear response functions and
correlation functions \cite{CKP97,ZBCK05}. There are two basic
approaches: static response theory relating equal time
fluctuations with infinite-time response and time-dependent
methods. Considering a number of different static response
relations, it was found that multiple definitions of effective
temperature converge to the same value within the context of an
athermal model of foam \cite{OODLLN02}. In contrast, the use of
time-dependent relations have revealed the impact of multiple time
scales on the dynamics. In simulations of driven, thermal systems,
it has been shown that the system exhibits the real temperature at
short times, but it has a higher effective temperature on long
time scales \cite{BB02}. In this case, the static effective
temperature and time-dependent effective temperature do not (and
can not) agree with each other. A recent surprising result is that
when considering {\it different} conjugate pairs of variables, the
static effective temperature determined from one pair can agree
with the time-dependent effective temperature from a different
pair \cite{OLN04}. The conditions for such agreement are not yet
understood, but recent simulation work for granular systems
suggests interesting directions of study \cite{PM06}.

Experiments probing fluctuation-dissipation measures in granular
systems have considered both diffusivities (a static measurement)
\cite{BU03} and the time-dependent fluctuation dissipation
relationship \cite{HM04}. The results on diffusivities is
consistent with simulations, at least in the sense that effective
temperatures from equipartition definitions do not tend to agree
with fluctuation-dissipation definitions if the inertial terms are
not relevant. The fluctuation-dissipation work considered a single
definition of temperature for a range of particle probe properties
\cite{HM04}, but it remains to compare different definitions of
effective temperature experimentally. Finally, an interesting
system that has been studied is ping-pong balls driven by flowing
air. In this case, the fluctuation-dissipation relation has been
confirmed \cite{OLDLD04}.

In this paper, we focus on a definition of effective temperature
using a time-dependent linear response measure and consider the
behavior as a function of density. The time dependent method
probes both the time-scale dependence of the effective temperature
and the {\it directional} dependence of the effective temperature.
The central issue is the behavior of the response function as a
function of the systems density. As the density increases, one
approaches the jamming transition. The jamming transition is the
point at which the system becomes ``frozen'' and can be
characterized by a diverging viscosity or the development of a
static elastic shear modulus for the system \cite{LN98,TPCSW01}.
Our results suggest that the jamming transition is a special point
where directional effects become important. Surprisingly, it is
also the point where time-dependent linear response measure is
most suggestive of a meaningful effective temperature.

The specific system reported on in this paper consists of plastic
spheres floating on the surface of water. The details of the
apparatus and measurement techniques are found in
Ref.~\cite{app,PD03}. The spheres are confined to the air-water
interface and placed between two concentric cylinders. A 50/50
mixture of spheres with radii of 0.32~cm and 0.24~cm is used. The
packing fraction $\rho$ is the total cross-sectional area of the
spheres at their midplane divided by the total area of the trough.
The radius of the outer cylinder is varied between 6~cm and
11.5~cm to vary the packing fraction. As jamming occurs with
increasing packing fraction, it is often useful to consider the
behavior in terms of the inverse of the packing fraction: $\phi =
1/\rho$. Measurements of the critical packing fraction for the
jamming of bead rafts have been previously reported \cite{TD05b}.
Based on these measurements, we focus on the range of packing
fractions close to, but below, the jamming transition. In this
regime, the system exhibits the properties of a power-law fluid
(viscosity proportional to the rate of strain to a power), and
there is no evidence of a yield-stress. Approaching the jamming
transition by varying the packing fraction at a fixed rate of
strain, the average stress, and by definition the viscosity,
diverge \cite{TD05b}.

We are interested in the behavior of the system with an imposed
constant rate of strain, or flow. Flow is generated by rotating
the outer cylinder. In order to measure the stress generated by
the flow, the inner cylinder is suspended on a torsion wire and
free to rotate. Angular displacements $\theta$ of the inner
cylinder are measured. The stress on the inner cylinder is
determined using $\sigma=\frac{\kappa \theta}{2 \pi r^{2}}$. Here
$\kappa$ is the torsion constant of the wire and $r = 2.7\ {\rm
cm}$ is the radius of the inner cylinder. Therefore, angular
fluctuations of the inner cylinder allow for measurement of stress
fluctuations. From a time series of the angular position of the
inner cylinder, we can measure correlation functions for either
the angular displacement of the inner cylinder or the stress
generated by the flow.

In addition to its use as a passive measuring device, an external
torque can be applied to the inner cylinder, and the corresponding
angular response measured. This is achieved through a coil of wire
that is attached to the inner cylinder and suspended within a
second coil, referred to as the outer coil. A current in the outer
coil generates a magnetic field, and by applying a voltage to the
inner coil to change its magnetic moment, a torque can be applied
to the inner coil \cite{app}. Because the position of the inner
cylinder is fluctuating during flow, in order to measure the
response function a repeated square wave pulse is applied to the
inner coil, and the response for each cycle is averaged. For each
measurement, the square pulse is applied several hundred times,
and the results averaged. An example of an average response for a
10~s on/10~s off pulse is shown in Fig.~1a. This represents a
period of 20~s. Figure 1b shows the individual response in the
case of pulses with a period of 50~s. To ensure that the system
has equilibrated after each on/off pulse, for the rest of the
paper all results used pulses with a 40~s period. This ensures
that turning on the pulse corresponds to applying a torque in the
direction of the external rotation, and because the system has
fully equilibrated, turning off the pulse is equivalent to
applying a torque opposite the direction of external rotation.

\begin{figure} [htb]
\begin{center}
\includegraphics[width=8cm]{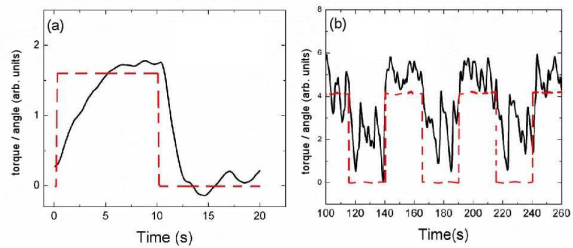}
\caption{(a) (color online) Averaged angular response (solid black
line) to an applied square-wave voltage pulse corresponding to an
applied torque (dashed red line). The angle and applied torque
have been scaled to appear on the same plot. (b) Time series of
angular response (solid black line) given an applied square-wave
voltage pulse that corresponds to an applied torque (dashed red
line). The angle and applied torque have been scaled to appear on
the same plot.}
\end{center}
\end{figure}

We will follow the conventions of Ref.~\cite{OLN04} and consider
the correlation function for the angular displacements of the
inner cylinder, $\tilde{C}_{\theta}(t)=<\theta(t)\theta(0)> -
<\theta>^2$, and the integrated response of $\theta$ to a small,
constant applied torque $\delta \tau$ given by
$\tilde{R}_{\theta}(t) = [<\theta(t)> - <\theta(0)>]/\delta \tau$.
Defining rescaled quantities,
$R_{\theta}(t)=\tilde{R}_{\theta}(t)/\tilde{C}_{\theta}(0)$ and
$C_{\theta}(t)=\tilde{C}_{\theta}(t)/\tilde{C}_{\theta}(0)$, the
fluctuation dissipation relation gives $R_{\theta}(t) = (1/T)(1 -
C_{\theta}(t))$, where $T$ is the temperature, or in our case, the
effective temperature. It is worth briefly considering the case of
an overdamped, torsional oscillator in a thermal bath described by
Langevin equation: $\kappa \theta + \alpha \dot{\theta} = \xi(t)$,
where $\kappa$ is the torsion constant, $\alpha$ is the damping
coefficient and $\xi(t)$ is the random force term for which
$<\xi(t)\xi(0)> = 2\alpha T \delta(t)$. For such a system,
$\tilde{C}_{\theta}(t) = (T/\kappa)\exp[-(\kappa/\alpha)t]$ and
$\tilde{R}_{\theta}(t) = (1/\kappa)(1 - exp[-(\kappa/\alpha)t)$.
From this, we see that the fluctuation dissipation relation is a
statement that the relaxation times for the correlation and
response function are the same.

\begin{figure} [htb]
\begin{center}
\includegraphics[width=8cm]{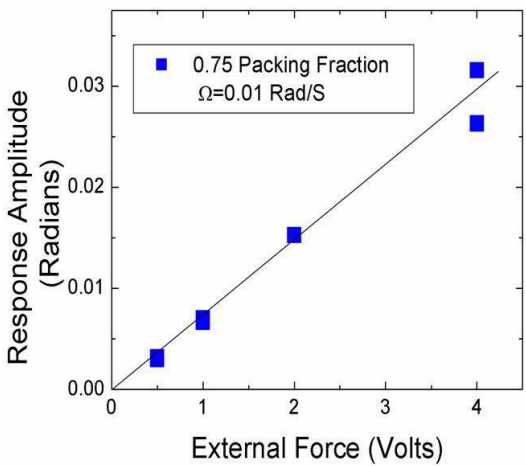}
\caption{The symbols are the final averaged angular displacement
of the inner cylinder as a function of the applied voltage pulse
given an external rotation rate of 0.01 rad/s and a packing
fraction of 0.75. The applied torque is linear in the applied
voltage.) }
\end{center}
\end{figure}

To test for linear response, we applied square waves of
increasingly larger amplitude voltages to the inner coil,
generating a torque on the inner cylinder. Figure 2 is a plot of
the final angular displacement of the inner cylinder versus
applied voltage for a packing fraction of 0.75 and a rotation rate
of 0.01 rad/s. This illustrates the linearity of the response. For
the range of voltages used, the response functions all collapsed
onto a single curve when scaled by the applied voltage, confirming
linear response. It should be noted that turning on the voltage
pulse corresponded to {\it decreasing} the applied rate of strain.
Likewise, turning off the pulse corresponded to {\it increasing}
the applied rate of strain.

\begin{figure} [htb]
\begin{center}
\includegraphics[width=8cm]{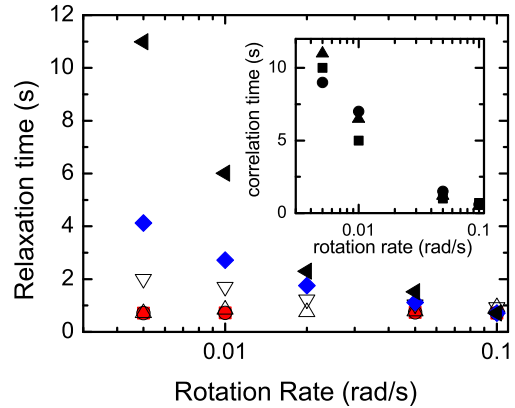}
\caption{(color online) Plot of the response times as a function
of rotation rate for three different packing fractions. Solid
symbols correspond to applying a torque in the direction of flow,
and open symbols correspond to a torque opposite the flow. The
insert is a plot of the correlation times as a function of
rotation rate for the same three packing fractions.}
\end{center}
\end{figure}

Before considering the connection between the response function
and the correlation function, it is useful to consider the time
scales associated with each separately. Figure~3 is a plot of the
relaxation times for different densities and rotation rates that
were obtained by fitting the response function to an exponential
curve. For comparison, the insert shows the correlation times for
the same densities as a function of rotation rate. It should be
noted that the correlation function in time was well fit by an
exponential (in frequency space, it was Lorentzian). This behavior
is consistent with modelling the inner cylinder as an over-damped
harmonic oscillator in a bath. The main feature of Fig.~3 is the
fact that the discrepancy between relaxation times for applied
torques parallel and anti-parallel to the applied rotation rate
develops only for high densities and slow rotation rates. For this
system, as the rotation rate is decreased, the average stress
during flow is decreased \cite{TD05b}. The jamming transition is
approached either by increasing the packing fraction or decreasing
the stress \cite{LN98}. Therefore, the anisotropy in response
develops as the jamming transition is approached.

\begin{figure} [htb]
\begin{center}
\includegraphics[width=8cm]{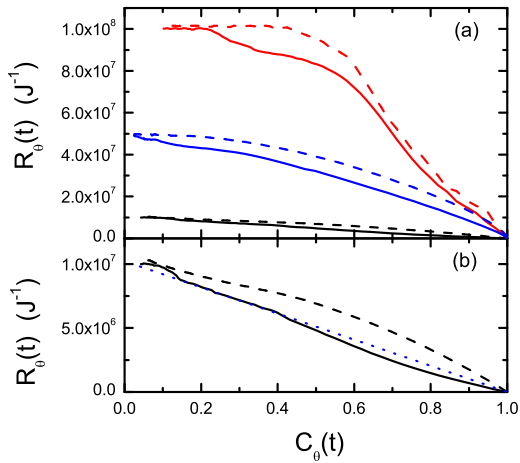}
\caption{(a) Parametric plots of the angular response function
versus the angular correlation function. The solid lines are for
the response when the applied torque is in the direction of flow.
The dashed lines are for the response when the applied torque is
in the opposite direction of the flow. In order of decreasing
amplitude for the response, the curves are for packing fractions
of 0.72, 0.74, and 0.78. (b) A close-up of the two curves for
packing fraction 0.78. The dashed and solid lines are the same
curves as in (a).}
\end{center}
\end{figure}

To highlight the features of the response and correlation
functions illustrated in Fig.~3, Fig.~4a consists of a series of
parametric plots of $R_{\theta}(t)$ as a function of
$C_{\theta}(t)$ for packing fractions, 0.72, 0.74, and 0.78,
respectively. The solid lines are for the response when the
applied torque is in the direction of flow (decreasing the rate of
strain). The dashed lines are for the response when the applied
torque is in the opposite direction as the rotation of the outer
cylinder (increasing rate of strain). If the system had a single
effective temperature, these plots would be straight lines with a
slope of $1/T$ and a y-intercept of $1/T$. Also, if the direction
of applied torque did not matter, the two curves for each density
would superimpose. As expected, as the density is increased, the
curves do not superimpose because the relaxation times are
different for the different directions of applied torques (see
Fig.~3).

In terms of defining an effective temperature, it is interesting
to note that as the packing fraction is increased, the response as
a function of the correlation approaches a straight line for the
case of a torque applied in the same direction as the flow. This
is illustrated in Fig.~4b where the results for a packing fraction
of 0.78 are plotted. The dotted line is a straight line guide to
the eye. In this case, it is interesting to compare the result
this gives for an effective temperature with other measures of
effective temperature to determine what, if any, significance
there is to the linear behavior. This will be the subject of
future studies.

The results reported here for the measured response functions
during the flow of plastic beads have a number of interesting
implications for generalizing fluctuation-dissipation relations as
methods of defining effective temperatures. First, they highlight
the fact that the applied rate of strain can break the symmetry of
the system and result in a directional dependence for the response
function. The existence of a directional dependence to the
response function is consistent with previous measurements of the
stress in granular systems in response to changes in applied
stress or rate of strain \cite{LBLG00}. These experiments focused
on the question of flow start up in an opposite direction to a
recently stopped applied flow. The initial applied rate of strain
forms stress chains with a directional preference. The difference
between flow start up in the direction and opposite the direction
of the initial flow can be understood in terms of the development
and breaking of these stress chains. This behavior occurs in the
context of a highly nonlinear change in rate of strain. In the
studies reported here, we observe similar directional dependence,
but in a regime where the applied torque is small enough that the
response is {\it linear} in the applied torque. An important
question is the behavior of the system as one continues to
decrease the driving amplitude. By definition, the truly linear
response function can not have the directional asymmetry. We are
currently working on increasing the resolution of the experiments
to determine if the true linear response function is
experimentally accessible. However, if the linear response regime
is not experimentally achievable, this will have important
implications for any practical definition of effective
temperature.

Finally, this asymmetry in the response function will not be a
generic feature of all definitions of effective temperature. For
example, experimental measurements in another granular system
probed the fluctuations perpendicular to the applied rate of
strain, where one would not expect anisotropic effects
\cite{HM04}. It will be important for future studies of effective
temperatures to determine when the directional asymmetry will
matter and how this impacts the generalization of definitions of
effective temperature in such cases.

\begin{acknowledgments}

This work was supported by Department of Energy grant
DE-FG02-03ED46071. The authors thank C. O'Hern, J. L. Barrat, L.
Bocquet, and P. Sollich for useful discussions.

\end{acknowledgments}


\begin{thebibliography}{23}
\expandafter\ifx\csname
natexlab\endcsname\relax\def\natexlab#1{#1}\fi
\expandafter\ifx\csname bibnamefont\endcsname\relax
  \def\bibnamefont#1{#1}\fi
\expandafter\ifx\csname bibfnamefont\endcsname\relax
  \def\bibfnamefont#1{#1}\fi
\expandafter\ifx\csname citenamefont\endcsname\relax
  \def\citenamefont#1{#1}\fi
\expandafter\ifx\csname url\endcsname\relax
  \def\url#1{\texttt{#1}}\fi
\expandafter\ifx\csname
urlprefix\endcsname\relax\def\urlprefix{URL }\fi
\providecommand{\bibinfo}[2]{#2}
\providecommand{\eprint}[2][]{\url{#2}}

\bibitem[{\citenamefont{Edwards and Makse}(2005)}]{EBM05}
\bibinfo{author}{\bibfnamefont{S.~F.} \bibnamefont{Edwards}} \bibnamefont{and}
  \bibinfo{author}{\bibfnamefont{J.~B. H.~A.} \bibnamefont{Makse}},
  \bibinfo{journal}{cond-mat} p. \bibinfo{pages}{0503057}
  (\bibinfo{year}{2005}).

\bibitem[{\citenamefont{Cugliandolo and Kurchan}(1999)}]{JC99}
\bibinfo{author}{\bibfnamefont{L.~F.} \bibnamefont{Cugliandolo}}
  \bibnamefont{and} \bibinfo{author}{\bibfnamefont{J.}~\bibnamefont{Kurchan}},
  \bibinfo{journal}{Physica A} \textbf{\bibinfo{volume}{263}},
  \bibinfo{pages}{242} (\bibinfo{year}{1999}).

\bibitem[{\citenamefont{Han and Grier}(2004)}]{DG04}
\bibinfo{author}{\bibfnamefont{Y.}~\bibnamefont{Han}} \bibnamefont{and}
  \bibinfo{author}{\bibfnamefont{D.~G.} \bibnamefont{Grier}},
  \bibinfo{journal}{Physical Review Letters} \textbf{\bibinfo{volume}{92}},
  \bibinfo{pages}{148301} (\bibinfo{year}{2004}).

\bibitem[{\citenamefont{Ciliberto et~al.}(2004)\citenamefont{Ciliberto,
  Garnier, Hernandez, Lacpatia, JF, and Chavarria}}]{CGHLPC04}
\bibinfo{author}{\bibfnamefont{S.}~\bibnamefont{Ciliberto}},
  \bibinfo{author}{\bibfnamefont{N.}~\bibnamefont{Garnier}},
  \bibinfo{author}{\bibfnamefont{S.}~\bibnamefont{Hernandez}},
  \bibinfo{author}{\bibfnamefont{C.}~\bibnamefont{Lacpatia}},
  \bibinfo{author}{\bibfnamefont{J.~F.~P.} \bibnamefont{JF}}, \bibnamefont{and}
  \bibinfo{author}{\bibfnamefont{G.~R.} \bibnamefont{Chavarria}},
  \bibinfo{journal}{Physica A} \textbf{\bibinfo{volume}{340}},
  \bibinfo{pages}{240} (\bibinfo{year}{2004}).

\bibitem[{\citenamefont{Goldburg et~al.}(2001)\citenamefont{Goldburg,
  Goldschmidt, and Kellay}}]{GGK01}
\bibinfo{author}{\bibfnamefont{W.~I.} \bibnamefont{Goldburg}},
  \bibinfo{author}{\bibfnamefont{Y.~Y.} \bibnamefont{Goldschmidt}},
  \bibnamefont{and} \bibinfo{author}{\bibfnamefont{H.}~\bibnamefont{Kellay}},
  \bibinfo{journal}{Phys. Rev. Lett} \textbf{\bibinfo{volume}{87}},
  \bibinfo{pages}{245502} (\bibinfo{year}{2001}).

\bibitem[{\citenamefont{Toth-Katona and Gleeson}(2003)}]{TG03}
\bibinfo{author}{\bibfnamefont{T.}~\bibnamefont{Toth-Katona}} \bibnamefont{and}
  \bibinfo{author}{\bibfnamefont{J.~T.} \bibnamefont{Gleeson}},
  \bibinfo{journal}{Phys. Rev. Lett} \textbf{\bibinfo{volume}{91}},
  \bibinfo{pages}{264501} (\bibinfo{year}{2003}).

\bibitem[{\citenamefont{D'Anna et~al.}(2003)\citenamefont{D'Anna, Mayor,
  Barrat, Loreto, and Nori}}]{GA03}
\bibinfo{author}{\bibfnamefont{G.}~\bibnamefont{D'Anna}},
  \bibinfo{author}{\bibfnamefont{P.}~\bibnamefont{Mayor}},
  \bibinfo{author}{\bibfnamefont{A.}~\bibnamefont{Barrat}},
  \bibinfo{author}{\bibfnamefont{V.}~\bibnamefont{Loreto}}, \bibnamefont{and}
  \bibinfo{author}{\bibfnamefont{F.}~\bibnamefont{Nori}},
  \bibinfo{journal}{Nature} \textbf{\bibinfo{volume}{424}},
  \bibinfo{pages}{909} (\bibinfo{year}{2003}).

\bibitem[{\citenamefont{Feitosa and Menon}(2004)}]{FM04}
\bibinfo{author}{\bibfnamefont{K.}~\bibnamefont{Feitosa}} \bibnamefont{and}
  \bibinfo{author}{\bibfnamefont{N.}~\bibnamefont{Menon}},
  \bibinfo{journal}{Phys. Rev. Lett} \textbf{\bibinfo{volume}{92}},
  \bibinfo{pages}{164301} (\bibinfo{year}{2004}).

\bibitem[{\citenamefont{Cugliandolo et~al.}(1997)\citenamefont{Cugliandolo,
  Kurchan, and Peliti}}]{CKP97}
\bibinfo{author}{\bibfnamefont{L.~F.} \bibnamefont{Cugliandolo}},
  \bibinfo{author}{\bibfnamefont{J.}~\bibnamefont{Kurchan}}, \bibnamefont{and}
  \bibinfo{author}{\bibfnamefont{L.}~\bibnamefont{Peliti}},
  \bibinfo{journal}{Phys. Rev. E} \textbf{\bibinfo{volume}{55}},
  \bibinfo{pages}{3898} (\bibinfo{year}{1997}).

\bibitem[{\citenamefont{Zamponi et~al.}(2005)\citenamefont{Zamponi, Bonetto,
  Cugliandolo, and Kurchan}}]{ZBCK05}
\bibinfo{author}{\bibfnamefont{F.}~\bibnamefont{Zamponi}},
  \bibinfo{author}{\bibfnamefont{F.}~\bibnamefont{Bonetto}},
  \bibinfo{author}{\bibfnamefont{L.~F.} \bibnamefont{Cugliandolo}},
  \bibnamefont{and} \bibinfo{author}{\bibfnamefont{J.}~\bibnamefont{Kurchan}},
  \bibinfo{journal}{J. Stat. Mech.: Theory and Experiment} p.
  \bibinfo{pages}{P09013} (\bibinfo{year}{2005}).

\bibitem[{\citenamefont{Ono et~al.}(2002)\citenamefont{Ono, O'Hern, Durian,
  Langer, Liu, and Nagel}}]{OODLLN02}
\bibinfo{author}{\bibfnamefont{I.~K.} \bibnamefont{Ono}},
  \bibinfo{author}{\bibfnamefont{C.~S.} \bibnamefont{O'Hern}},
  \bibinfo{author}{\bibfnamefont{D.~J.} \bibnamefont{Durian}},
  \bibinfo{author}{\bibfnamefont{S.~A.} \bibnamefont{Langer}},
  \bibinfo{author}{\bibfnamefont{A.~J.} \bibnamefont{Liu}}, \bibnamefont{and}
  \bibinfo{author}{\bibfnamefont{S.~R.} \bibnamefont{Nagel}},
  \bibinfo{journal}{Phys. Rev. Lett.} \textbf{\bibinfo{volume}{89}},
  \bibinfo{pages}{095703} (\bibinfo{year}{2002}).

\bibitem[{\citenamefont{Berthier and Barrat}(2002)}]{BB02}
\bibinfo{author}{\bibfnamefont{L.}~\bibnamefont{Berthier}} \bibnamefont{and}
  \bibinfo{author}{\bibfnamefont{J.-L.} \bibnamefont{Barrat}},
  \bibinfo{journal}{Phys. Rev. Lett.} \textbf{\bibinfo{volume}{89}},
  \bibinfo{pages}{095702} (\bibinfo{year}{2002}).

\bibitem[{\citenamefont{O'Hern et~al.}(2004)\citenamefont{O'Hern, Liu, and
  Nagel}}]{OLN04}
\bibinfo{author}{\bibfnamefont{C.~S.} \bibnamefont{O'Hern}},
  \bibinfo{author}{\bibfnamefont{A.~J.} \bibnamefont{Liu}}, \bibnamefont{and}
  \bibinfo{author}{\bibfnamefont{S.~R.} \bibnamefont{Nagel}},
  \bibinfo{journal}{Phys. Rev. Lett.} \textbf{\bibinfo{volume}{93}},
  \bibinfo{pages}{165702} (\bibinfo{year}{2004}).

\bibitem[{\citenamefont{Potiquar and Makse}(2006)}]{PM06}
\bibinfo{author}{\bibfnamefont{F.~Q.} \bibnamefont{Potiquar}} \bibnamefont{and}
  \bibinfo{author}{\bibfnamefont{H.~A.} \bibnamefont{Makse}},
  \bibinfo{journal}{Euro. Phys. Jour. E} \textbf{\bibinfo{volume}{19}},
  \bibinfo{pages}{171} (\bibinfo{year}{2006}).

\bibitem[{\citenamefont{Utter and Behringer}(2004)}]{BU03}
\bibinfo{author}{\bibfnamefont{B.}~\bibnamefont{Utter}} \bibnamefont{and}
  \bibinfo{author}{\bibfnamefont{R.~P.} \bibnamefont{Behringer}},
  \bibinfo{journal}{Physical Review E} \textbf{\bibinfo{volume}{69}},
  \bibinfo{pages}{031308} (\bibinfo{year}{2004}).

\bibitem[{\citenamefont{Song et~al.}(2005)\citenamefont{Song, Wang, and
  Makse}}]{HM04}
\bibinfo{author}{\bibfnamefont{C.}~\bibnamefont{Song}},
  \bibinfo{author}{\bibfnamefont{P.}~\bibnamefont{Wang}}, \bibnamefont{and}
  \bibinfo{author}{\bibfnamefont{H.}~\bibnamefont{Makse}},
  \bibinfo{journal}{Proc. Nat. Academy} \textbf{\bibinfo{volume}{102}},
  \bibinfo{pages}{2299} (\bibinfo{year}{2005}).

\bibitem[{\citenamefont{Ojha et~al.}(2004)\citenamefont{Ojha, Lemieux, Dixon,
  Liu, and Durian}}]{OLDLD04}
\bibinfo{author}{\bibfnamefont{R.}~\bibnamefont{Ojha}},
  \bibinfo{author}{\bibfnamefont{P.-A.} \bibnamefont{Lemieux}},
  \bibinfo{author}{\bibfnamefont{P.}~\bibnamefont{Dixon}},
  \bibinfo{author}{\bibfnamefont{A.}~\bibnamefont{Liu}}, \bibnamefont{and}
  \bibinfo{author}{\bibfnamefont{D.}~\bibnamefont{Durian}},
  \bibinfo{journal}{Nature} \textbf{\bibinfo{volume}{427}},
  \bibinfo{pages}{521} (\bibinfo{year}{2004}).

\bibitem[{\citenamefont{Liu and Nagel}(1998)}]{LN98}
\bibinfo{author}{\bibfnamefont{A.~J.} \bibnamefont{Liu}} \bibnamefont{and}
  \bibinfo{author}{\bibfnamefont{S.~R.} \bibnamefont{Nagel}},
  \bibinfo{journal}{Nature} \textbf{\bibinfo{volume}{396}}, \bibinfo{pages}{21}
  (\bibinfo{year}{1998}).

\bibitem[{\citenamefont{Trappe et~al.}(2001)\citenamefont{Trappe, Prasad,
  Cipelletti, Segre, and Weitz}}]{TPCSW01}
\bibinfo{author}{\bibfnamefont{V.}~\bibnamefont{Trappe}},
  \bibinfo{author}{\bibfnamefont{V.}~\bibnamefont{Prasad}},
  \bibinfo{author}{\bibfnamefont{L.}~\bibnamefont{Cipelletti}},
  \bibinfo{author}{\bibfnamefont{P.~N.} \bibnamefont{Segre}}, \bibnamefont{and}
  \bibinfo{author}{\bibfnamefont{D.~A.} \bibnamefont{Weitz}},
  \bibinfo{journal}{Nature} \textbf{\bibinfo{volume}{411}},
  \bibinfo{pages}{772} (\bibinfo{year}{2001}).

\bibitem[{\citenamefont{Ghaskadvi and Dennin}(1998)}]{app}
\bibinfo{author}{\bibfnamefont{R.~S.} \bibnamefont{Ghaskadvi}}
  \bibnamefont{and} \bibinfo{author}{\bibfnamefont{M.}~\bibnamefont{Dennin}},
  \bibinfo{journal}{Rev. Sci. Instr.} \textbf{\bibinfo{volume}{69}},
  \bibinfo{pages}{3568} (\bibinfo{year}{1998}).

\bibitem[{\citenamefont{Pratt and Dennin}(2003)}]{PD03}
\bibinfo{author}{\bibfnamefont{E.}~\bibnamefont{Pratt}} \bibnamefont{and}
  \bibinfo{author}{\bibfnamefont{M.}~\bibnamefont{Dennin}},
  \bibinfo{journal}{Phys. Rev. E} \textbf{\bibinfo{volume}{67}},
  \bibinfo{pages}{051402} (\bibinfo{year}{2003}).

\bibitem[{\citenamefont{Twardos and Dennin}(2005)}]{TD05b}
\bibinfo{author}{\bibfnamefont{M.}~\bibnamefont{Twardos}} \bibnamefont{and}
  \bibinfo{author}{\bibfnamefont{M.}~\bibnamefont{Dennin}},
  \bibinfo{journal}{Granular Matter} \textbf{\bibinfo{volume}{7}},
  \bibinfo{pages}{92} (\bibinfo{year}{2005}).

\bibitem[{\citenamefont{Losert et~al.}(2000)\citenamefont{Losert, Bocquet,
  Lubensky, and Gollub}}]{LBLG00}
\bibinfo{author}{\bibfnamefont{W.}~\bibnamefont{Losert}},
  \bibinfo{author}{\bibfnamefont{L.}~\bibnamefont{Bocquet}},
  \bibinfo{author}{\bibfnamefont{T.~C.} \bibnamefont{Lubensky}},
  \bibnamefont{and} \bibinfo{author}{\bibfnamefont{J.~P.}
  \bibnamefont{Gollub}}, \bibinfo{journal}{Phys. Rev. Lett.}
  \textbf{\bibinfo{volume}{85}}, \bibinfo{pages}{1428} (\bibinfo{year}{2000}).

\end{thebibliography}

\end{document}